\documentclass[12pt]{iopart}
\usepackage{setstack,latexsym,float,color}
\usepackage{graphicx}

\begin{document}

\title[Single$-$variant bottleneck in bacteremia]{Single variant
  bottleneck in the early dynamics of {\em H.~influenzae} bacteremia
  in neonatal rats questions the theory of independent action}

\author{Xinxian Shao$^1$, Bruce Levin$^2$,
Ilya Nemenman$^{1,2}$}

\address{$^1$ Department of Physics, Emory University, Atlanta, GA 30322, USA}
\address{$^2$ Department of Biology, Emory University, Atlanta, GA 30322, USA}
\ead{ilya.nemenman@emory.edu}
\vspace{10pt}

\begin{abstract}
  There is an abundance of information about the genetic basis,
  physiological and molecular mechanisms of bacterial pathogenesis. In
  contrast, relatively little is known about population dynamic
  processes, by which bacteria colonize hosts and invade tissues and
  cells and thereby cause disease. In an article published in 1978,
  Moxon and Murphy presented evidence that, when inoculated
  intranasally with a mixture streptomycin sensitive and resistant
  (Sm$^S$ and Sm$^R$) and otherwise isogenic stains of
  \emph{Haemophilus influenzae} type b (\emph{Hib}), neonatal rats
  develop a bacteremic infection that often is dominated by only one
  strain, Sm$^S$ or Sm$^R$. After rulling out other possibilities
  through years of related experiments, the field seems to have
  settled on a plausible explanation for this phenomenon: the first
  bacterium to invade the host activates the host immune response that
  `shuts the door' on the second invading strain. To explore this
  hypothesis in a necessarily quantitative way, we modeled this
  process with a set of mixed stochastic and deterministic
  differential equations.  Our analysis of the properties of this
  model with realistic parameters suggests that this hypothesis cannot
  explain the experimental results of Moxon and Murphy, and in
  particular the observed relationship between the frequency of
  different types of blood infections (bacteremias) and the inoculum
  size. We propose modifications to the model that come closer to
  explaining these data. However, the modified and better fitting
  model contradicts the common theory of independent action of
  individual bacteria in establishing infections. We discuss the
  implications of these results.
  
\end{abstract}

%
%
\submitto{\PB}
%
%
%

\section{Introduction}
Before the {\em Hib} ({\em Haemophilus influenzae} type b) conjugate
vaccine was developed and taken into routine use in the U.\ S., {\em
  H.~influenzae} was the leading cause of bacterial meningitis in
children under the age of five \cite{thigpen2011bacterial}. At the
same time, bacterial meningitis had high mortality and serious
sequela, including deafness, blindness and mental retardation. Even
today, at least in part because of the lack of vaccines, in the
developing world the mortality rate from {\em H. influenzae }
infections is substantial, with case mortalities approaching 14.3\%
\cite{thigpen2011bacterial}.

\begin{figure}
\centering
    \includegraphics[width=.75\linewidth]{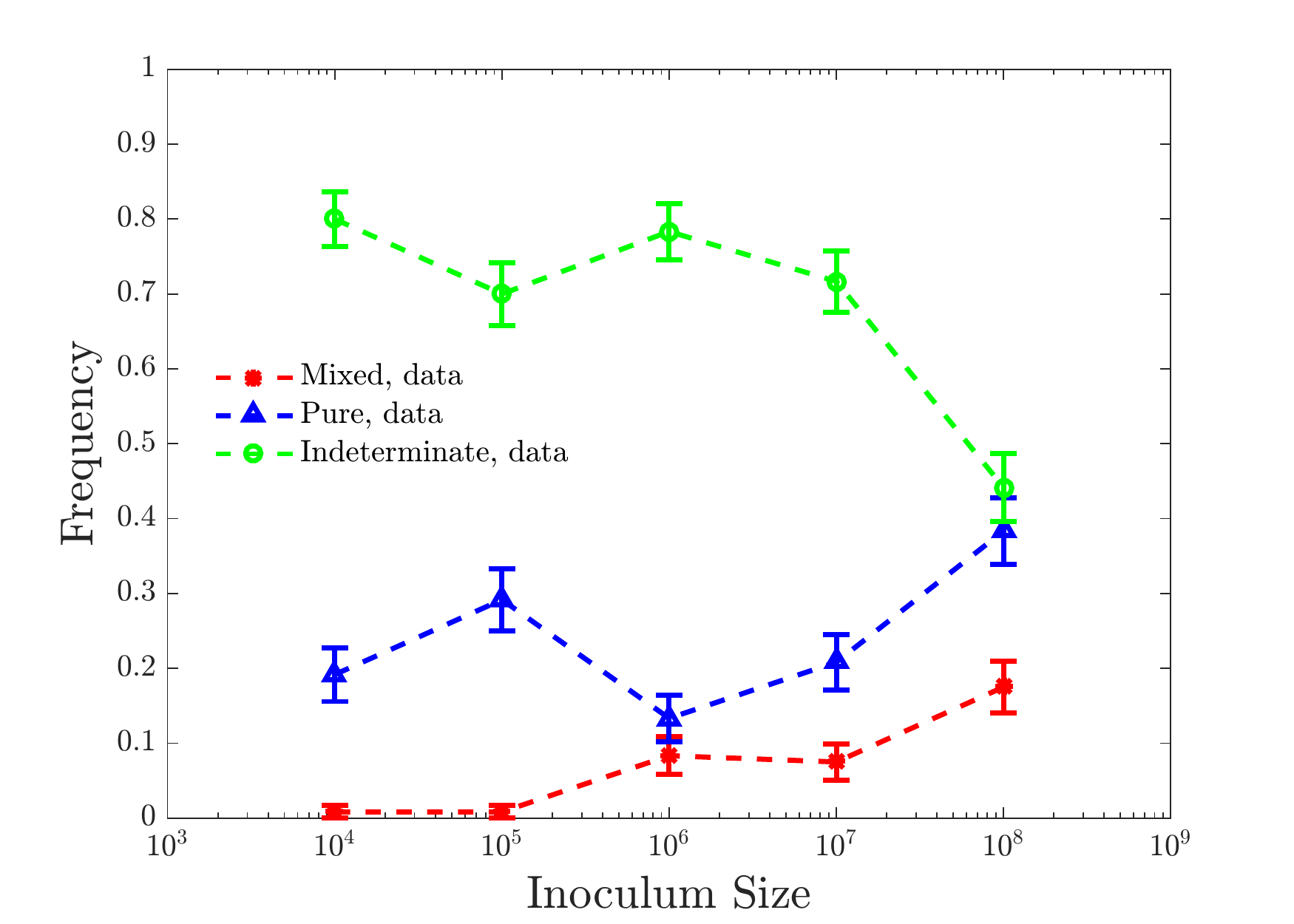}
  \caption
  {{\bf Moxon and Murphy's experimental data.}  Replotted from
    \cite{MoxonMurphy78}.  120 neonatal rats were infected at each
    inoculum size, which ranged from $10^4$ to $10^8$ bacteria,
    equally mixed from streptomycine susceptible strain, Sm$^S$, and
    streptomycine resistant strain, Sm$^R$.  Blood of the rats was
    then harvested and cultured.  All cultures that produced both
    Sm$^S$ and Sm$^R$ colonies were called {\em mixed}
    infections. All cultures that produced 8 or more colonies of one
    strain and none of the other were called {\em pure} infections
    (there was no statistically significant difference in the
    abundance of pure Sm$^S$ or pure Sm$^R$ cultures). Cultures that
    produced no colonieshown on the 
    plots, or produced colonies of one strain only, but
    fewer than 8 of those, were called indeterminate. Most infections
    ended up as pure (single-variant) infections 54 hours
    post-inoculum.  Samples taken within 5 and 30 minutes
    post-inoculum were mixed (data not show on the plot), see
    Ref.~\cite{MoxonMurphy78}). Error bars denote the usual
    square-root counting errors. Over four orders of magnitude in the
    inoculum size, preponderance of indeterminate infections declined
    somewhat, from $\sim$80\% to $\sim$40\%.  Over the same range of
    the inoculum, mixed infections increased from $<$1\% to
    $\sim$20\%. }
\label{fig:mmdata}
\end{figure}

In 1974 Richard Moxon and colleagues developed a neonatal rat model to
study the pathogenesis of \emph{Haemophilus influenzae}
\cite{moxon1974haemophilus}. They presented evidence that {\em Hib}
infection could be divided into three elements: nasopharyngeal
colonization, bacteremia, and central nervous system (CNS)
invasion. In 1978, by intranasally inoculating the neonatal rats with
mixtures of otherwise isogenic streptomycin sensitive and resistant
strains of {\em H.~influenzae}, Sm$^S$ and Sm$^R$, and tracking the
development of bacteremia and meningitis, Moxon and Murphy found that,
five minutes after inoculation, {\em both variants} were found in the
blood.  In contrast at 54 hours, nearly 70\% of rats had {\em pure} Sm$^S$ or
Sm$^R$ in the blood (Figure \ref{fig:mmdata}) and cerebrospinal
fluid, and the cultures had Sm$^S$ and Sm$^R$ isolated in nearly
equal frequency \cite{MoxonMurphy78}. Their observation that the
primary infections (nasal colonization and early blood flora for {\em
  Hib}) are diverse, while mature infections (blood after $>10$ hrs
for {\em Hib}) were monoclonal is known as the {\em single-variant
  bottleneck}. The bottleneck is not unique to {\em Hib}. In fact, it
was discovered first by Meynell in {\em Salmonella typhimurium}
infections in mice \cite{meynell1957some}. And similar observations
have been reported in experimental studies of other host-bacteria
\cite{abel2015sequence,kaiser2013lymph,abel2015analysis} and
host-virus
\cite{liu1951studies,donald1954counts,liu1953studies,fischer2010transmission,
  bull2011sequential,haaland2009inflammatory,wang2010hepatitis}
interactions.

Models of the infection process must be able to explain the
single-variant bottleneck, and also the fact that, although there are
high frequencies of people colonized with virulent strains of
\emph{H.~influenzae} as carriers, only minority of hosts (e.~g.,
children under 5 \cite{MoxonMurphy78,thigpen2011bacterial}) manifest
invasive infections even for the most virulent strains of these
bacteria.  Three classes of possible explanations have been proposed:
stochasticity resulting from {\em independent action} of bacteria
\cite{meynell1957some,meynell1957applicability,rubin1987bacterial,
  balaban2004bacterial,kussell2005phenotypic}, variations in host
susceptibility, and the emergence of bacterial mutants (known as
within-host evolution) \cite{levin2001we,meyers2003epidemiology,
  margolis2007within}. The original theory of independent action was
proposed by Druett in 1952 \cite{druett1952bacterial}. It assumed that
each individual bacterial cell has an independent probability to
colonize the host, and that an infection can start from a single
random bacterium, hence explaining the bottleneck. Meynell and Stocker
later provided experimental evidence consistent with the theory of
independent action \cite{meynell1957some} and inconsistent with an
alternative: synergistic or cooperative action.  However, in its
simple form, independent action could not explain why a variant
present in the blood five minutes post-inoculation can no longer be
detected a few hours later. Host susceptibility could only explain the
high infection rate in young children over adults, but not the
experimental observation of the presence of both variants in the
bloodstream only very early post-inoculation stage
\cite{MoxonMurphy78}. Finally, to test the within-host evolution
hypothesis, Margolis and Levin performed additional {\em
  H.~influenzae} experiments with neonatal rats. They compared the
invasiveness and the re-colonization potential of the variant
surviving in the bloodstream and the remaining variant staying in
nasopharynx \cite{margolis2007within}. In five out of six clones
examined, they observed no difference in the ability of bacteria
isolated from the blood to outcompete those remaining in the
nasopharynx during re-infection. This result suggested that
within-host evolution is not the dominant explanation for the
Moxon-Murphy observations.

During the last two decades it has become clear that bacteria can
switch phenotypic states without any changes in their genomes
\cite{meynell1957applicability,balaban2004bacterial,kaufmann2007heritable,
  ackermann2008self,acar2008stochastic,elowitz2002stochastic,
  kaern2005stochasticity}.  This provides an alternative hypothesis to
explain the Moxon-Murphy results. First, a single bacterium of one
strain can randomly switch into a faster growing phenotype or a more
invasive state.  Then the host immune system, activated by the
infection emerging from the switched bacterium, acts against both
strains and clears the clone that has not phenotypically switched and
thus grows slowly.  In other words, the first successfully switching
cell in one strain will interact with the host immune response to
`shut the door' on the other strain and thereby explain the
single-variant bottleneck and the subsequent failure to isolate
genetically more invasive clones from the blood
\cite{margolis2007within}.

This stochastic switching mechanism together with the immune response
has been mentioned frequently as a possible explanation for the
bottleneck phenomenon in various presentations and discussions.
Nevertheless, we have been unable to find a detailed theoretical or
experimental analysis of this process in the literature. We call this
phenotypic switching mechanism followed by immune-facilitated
clearance of the competing strain the {\em colloquial hypothesis}.
Our goal here is to analyze the colloquial hypothesis quantitatively
and to explore the condition under which it can or cannot rescue the
theory of independent action as the explanation of the single-variant
bottleneck in early bacteremia.

In the following sections, we will develop a mathematical model of the
colloquial hypothesis applied to the early stages of {\em Hib}
bacteremia inoculated with two variants of equally invasive bacteria.
We will show that the model, as well as its simple extensions, cannot
provide a quantitative explanation for the experimental data
\cite{MoxonMurphy78,meyers2003epidemiology,margolis2007within,moxon1994adaptive},
and namely the observed weak dependence of the rate of different types
of infections on the inoculum size. We argue that, to provide even a
semi-quantitative fit to the data, we must assume that various rate
parameters describing the infection scale {\em sublinearly} with the
inoculum size, so that the probability per bacterium to start an
infection decreases when other bacteria are present. This means
further evidence for abandoning the theory of independent action.

\section{Hypothesis and Model} 

Inspired by demonstration of phenotypic switching in bacteria
\cite{balaban2004bacterial,ackermann2008self,acar2008stochastic,
  elowitz2002stochastic,kaern2005stochasticity,dubnau2006bistability,
  davidson2008individuality,jarboe2004stochastic}, we propose that
each individual bacterial cell of both variants A and B has two
phenotypes relevant for the early infection. The first is the
``crossing'' phenotype (C), which allows bacteria to cross the
physical barrier between the nasopharynx and blood, but does not
exhibit strong growth in the bloodstream \cite{o1985gal,
  pelevs2006reduction,casadesus2013pap}. The second is the ``growing''
phenotype (G), with cells that grow fast in the bloodstream, but
cannot cross into the bloodstream.  After a bacterium crosses into the
bloodstream, it can switch to the G state, but the switching C$\to$G
is stochastic and rare. In this work, we are not concerned with the
mechanisms underlying the existence of these two states and of
switching between them, but only focus on consequences of the
switching.

Once bacterial cells enter the bloodstream, immune response is
activated. To model the immune response in the early stages of
bacteremia, we assume that neonatal rats only have innate immunity,
which is non-specific and responds as soon as the bacterial cells
emerge in the blood
\cite{thigpen2011bacterial,MoxonMurphy78,levin2001we,antia1994model}. In
other words, there is no clonal expansion, and instead there is a
finite reservoir of immune cells that can be recruited to the
infection site linearly until the reservoir is depleted
\cite{antia1994model}.

These assumptions are represented in the following ordinary
differential equations (ODEs) describing the growth of bacterial cells
of variant A and the immune cell recruitment:
\begin{eqnarray}
\label{eq:AC}
\frac{dA_C}{dt}&=& g_CA_C \left(1-\frac{N_{\rm total}}{N_0}
                   \right)-\gamma_CIA_C- dA_C +j, \\
\label{eq:AG}
\frac{dA_G}{dt}&=&g_GA_G \left(1-\frac{N_{\rm total}}{N_0} \right)-\gamma_GIA_G-dA_G,\\
\label{eq:Im}
\frac{dI}{dt}& =&r_C (I_0-I)(A_C+B_C)+r_G(I_0-I)(A_G +B_G)-d_II . 
\end{eqnarray}
The growth of the bacterial strain B is described by equations similar
to Eqs.~(\ref{eq:AC}, \ref{eq:AG}), with letters A replaced by B.

In the equations describing bacterial population dynamics,
Eqs.~(\ref{eq:AC}, \ref{eq:AG}), $g_C$ and $g_G$ are the growth rates
of the crossing and the growing phenotypes, respectively (same for
variants A and B since both variants are equally virulent
\cite{MoxonMurphy78, margolis2007within}). For simplicity, in what
follows we set $g_C=0$.  $N_{\rm total}$ is the total number of
bacteria in the blood, $N_{\rm total}\equiv A_C+A_G+B_C+B_G$.  $N_0$
is the carrying capacity, the maximum of the bacterial population in
the bloodstream, so that $N_0\ge N_{\rm total}$.  $\gamma_C$ and $\gamma_G$ are
the bacterial death rates due to the elimination by the immune cells,
and $d$ is the natural cell death. Finally, $j$ is the flux of the
crossing phenotype cells from the nasopharynx to the bloodstream per
unit time. To satisfy the hypothesis of independent action, it is
assumed to be linearly proportional to the inoculum size $S$, so that
$j =\alpha_j S$, where $\alpha_j$ is some constant.

Equation (\ref{eq:Im}) describes the immune cells recruitment. Here
$r_C$ and $r_G$ are the recruitment rates due to effects of
the bacterial phenotypes \cite{antia1994model}. 
$I_0$ is the total number of available
innate immune cells in the host. $d_I$ is the death rate, or deactivation 
rate of immune cells. 
Parameters in Eq.~(\ref{eq:Im}) are determined up to a scale. Thus we
set $I_0=1$, which redefines the scale of $I$ and also renormalizes
$\gamma_C$ and $\gamma_G$ in Eqs.~(\ref{eq:AC}, \ref{eq:AG}). 
To simplify the model,
we set $\gamma_C=\gamma_G=\gamma$, and $r_C=r_G=r$.
  
To finish specifying the model, we assume for now that the switching
from C to G is a single step stochastic transition at a low
per-bacterium rate $\rho$. Since the switching is single-step, the
waiting time to the switch is exponentially distributed for each
cell. Further, if the independent action hypothesis is valid, then for
$A_C$ bacteria in the crossing phenotype, the probability of having
$k$ individuals of type A switching to the growing state per time
$\Delta t$ is given by the Poisson distribution:
\begin{eqnarray}
\label{eq:ppoiss}
 P(k|A_C)=\frac{(\rho A_C\Delta t)^ke^{-\rho A_C\Delta t}}{k!}.
\end{eqnarray}
A similar distribution determines the switching probability for the B
strain. We do not consider switching back from the G to the C state.

We simulate the model using the Euler method to solve its ODEs,
Eqs.~(\ref{eq:AC}, \ref{eq:AG}, \ref{eq:Im}) and their equivalents for
the B strain. Further, at every time step, we generate a random number
of switching individuals using Eq.~(\ref{eq:ppoiss}) for the A and the
B strain. If the number of cells in any strain / phenotypic state
combination falls below one, we set it to zero to account for the
discreteness of the bacteria. While this combined
stochastic-deterministic simulation scheme is certainly not the most
accurate, we feel that it offers the precision necessary for our
analysis. It certainly is capable of discovering the salient
qualitative features of the models that we investigate. Further, it is
much faster that fully stochastic simulations schemes, which is
important since the model must be solved repeatedly in optimization
steps. Finally, statistical properties of {\em Hib} bacterial division
are not well understood; thus building a fully stochastic model of the
system, including modeling the divisions as first order, memoryless
reactions, would be not any more accurate than neglecting the
stochasticity altogether.

\section{Results} 

\subsection{The colloquial model}

\begin{figure} 
\centering
    \includegraphics[width=1\linewidth]{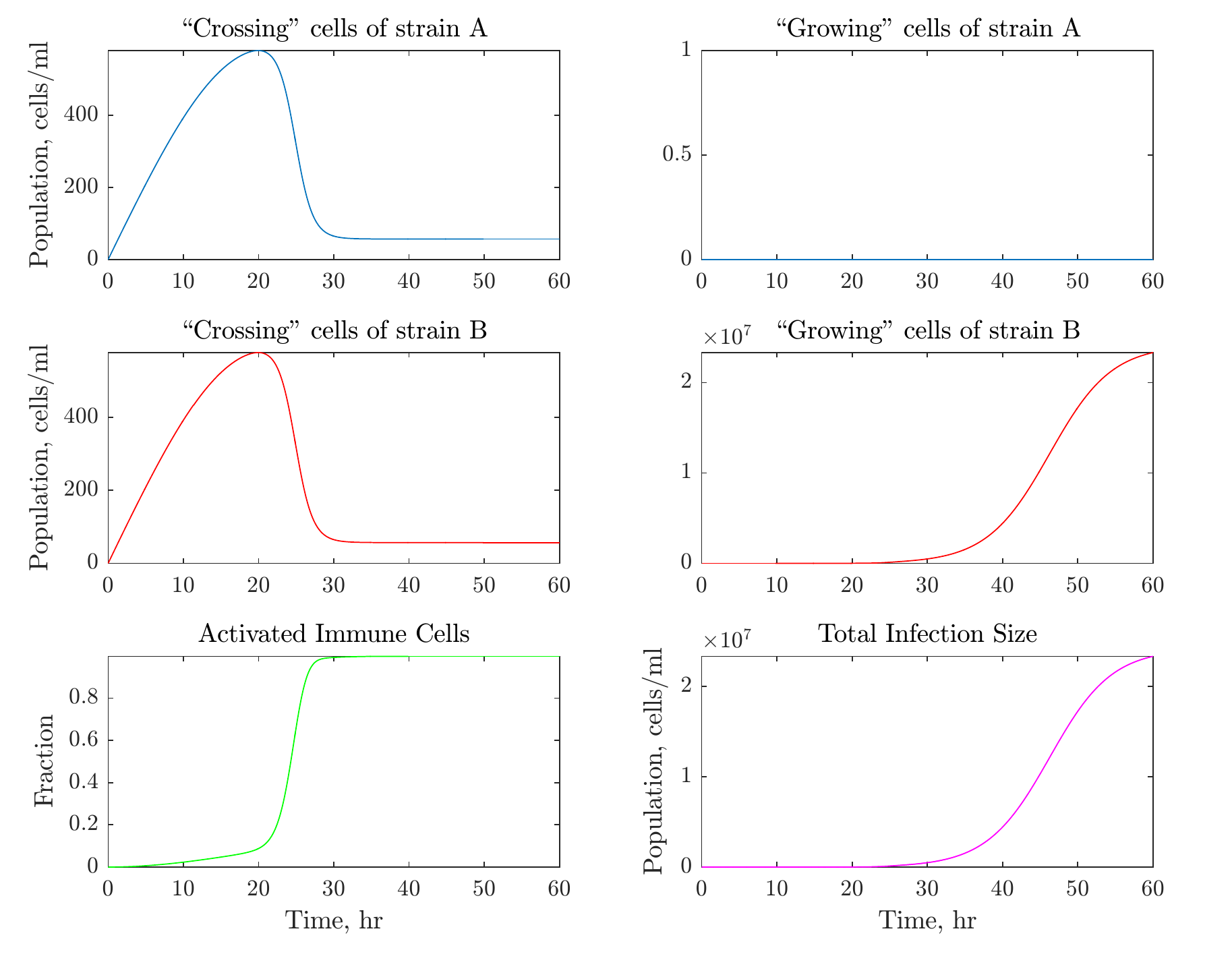}
    \caption{{\bf Simulation of early bacteremia resulting in a pure
        infection in an individual rat.} Simulations were done with
      the following parameters: inoculum size $S=10^{5}$; switching
      rate $\rho=4.5\cdot 10^{-5}$ h$^{-1}$; growth rates $g_C=0$,
      $g_G=1$ h$^{-1}$; immune recruitment rates $r=6.1\cdot 10^{-6}$
      h$^{-1}$; rate at which immune cells kill bacteria $\gamma=0.75$
      h$^{-1}$; carrying capacity of the blood $N_0=10^{8}$ cells;
      flux from the nasopharynx to blood $\alpha_j=4.3\cdot10^{-4}$
      h$^{-1}$; natural death rate of bacteria and immune cells
      $d = 0.01$ h$^{-1}$ and $d_I = 0.02$ h$^{-1}$. In this
      realization, variant $B$ has the first switch from $C$ to $G$
      at about 11 hours and establishes bacteremia. The panels show
      the population size of the crossing and the growing phenotypes
      of A and B strains, the fraction of the immune response
      activated, and the total infection size. Notice that the
      vertical axes in different panels have different scalings.}
\label{mmpureb}
\end{figure}

We illustrate a possible dynamics of the colloquial model in
Fig.~\ref{mmpureb} for the first 60 hours post-inoculum in an
individual rat with a certain set of model parameters. In this case, a
cell of strain B switched to the growing phenotype first at
$t\approx11$ hrs.  Then the rapid growth of $B_G$ accelerated the
recruitment of immune cells. In their turn, the immune cells nearly
wiped the population of the non-switched strain A, transforming the
infection into the pure B infection about 30 h
post-inoculation. Therefore, even though cells act independently, they
interact through the immune response, and the first variant to have a
switcher wins the competition. This example illustrates that the
colloquial hypothesis may have a potential to explain the
single-variant bottleneck in the early stages of bacteremia.

To test the suitability of the colloquial model quantitatively, we
calculate and maximize its likelihood given the observed experimental
data. As in the Moxon and Murphy experiment, we assume the multinomial
structure of the data with three possible outcomes: pure infection,
mixed infection, and indeterminate infection. Recall that Moxon and
Murphy plated blood samples from their rats and counted the number of
colonies of each strain that grew as a result. They defined any
culture with colonies of both strains (even if one of the strains had
as few as one colony) as a mixed infection. A pure infection was
defined more stringently, so that there had to be at least eight
colonies of one strain and none of the other to qualify. All other
cases were deemed indeterminate. In our simulations, accounting for
dilution at plating, we estimate that a mixed infection would require
both bacterial strains present at a level of 100 cells/ml of bacteria
or more, and a pure infection would require at least 800 cells/ml of
bacteria of one type and less than 100 cells/ml of the other. To
calculate the likelihood of the data given a set of parameters, we
simulate infections using our mixed stochastic-deterministic
simulations. We repeat this 200 times to estimate the multinomial
probabilities. We then write down the multinomial likelihood of the
experimental data given the frequencies defined by the numerical
simulations. Finally, we optimize the model over the parameters using
{\tt patternsearch} from Matlab with GPS Positive basis Np1 as the
poll method. This routine allows optimization of stochastic
functions. The optimization is performed at least three times from
different initial conditions, and we report the best fit model as the
one maximizing the likelihood over all such optimization runs.

The experimental data that we fit contains five different inoculum
sizes, and three possible outcomes at 54 hrs post inoculation (for a
total of 10 independent data points since the frequencies at each
inoculum sum to one). In addition, the experimental data contains
measurements a few minutes after inoculation for each inoculum size
(10 more independent data points), at which point {\em every}
infection was mixed. Note that these data are not time series---every
rat could be analyzed only once---so that the data at
different time points are independently multinomially distributed.  

These 20 data points must be explained by 8 independent parameters:
$\alpha_j$, $g_G$, $d$, $N_0$, $r$, $\gamma$, $d_I$, and $\rho$. This
may sound like an easy fitting problem. However, it turns out that the
requirement of having all mixed infections soon after inoculation, and
a lot of pure infections later on is not easy to satisfy.  Thus we do
not perform formal analysis of the quality of fit / overfitting in
this and the other models we try: the difficulty to fit the data makes
most models obviously poor, and differences among the quality of fits
of various models are clear without formal analyses.

The optimization is further constrained since biologically realistic
limits exist on the model parameters. First, all of the parameters are
positive. Further, an upper limit on $N_0$ is about
$\sim 2\cdot10^{9}$ cells/ml \cite{artman1983growth}, while its lower
limit is determined by the fact that Moxon and Murphy observed
$\sim 1\times10^{4}$ or more cells/ml in the bloodstream of neonatal
rats with severe infections \cite{MoxonMurphy78}.  The growth rate of
{\em Hib} in synthetic blood culture was studied in
\cite{artman1983growth}, which provides the initial guess and upper
limit of $g_G$ between 0.4 to 1.2 per hour.  We could not find any
data in the literature regarding the parameters of the innate immune
response to {\em Hib}.  However, some data is available for {\em
  Listeria} infection \cite{bancroft1986regulation,bancroft1987cell},
which allowed us to choose initial conditions of the immune response
for the optimization: $r\sim 1\cdot10^{-6}$ h$^{-1}$ cells$^{-1}$,
$\gamma\sim0.1$ h$^{-1}$, $d_I\sim 0.02$ h$^{-1}$.

\begin{figure}[h]
\centering
    \includegraphics[width=.5\linewidth]{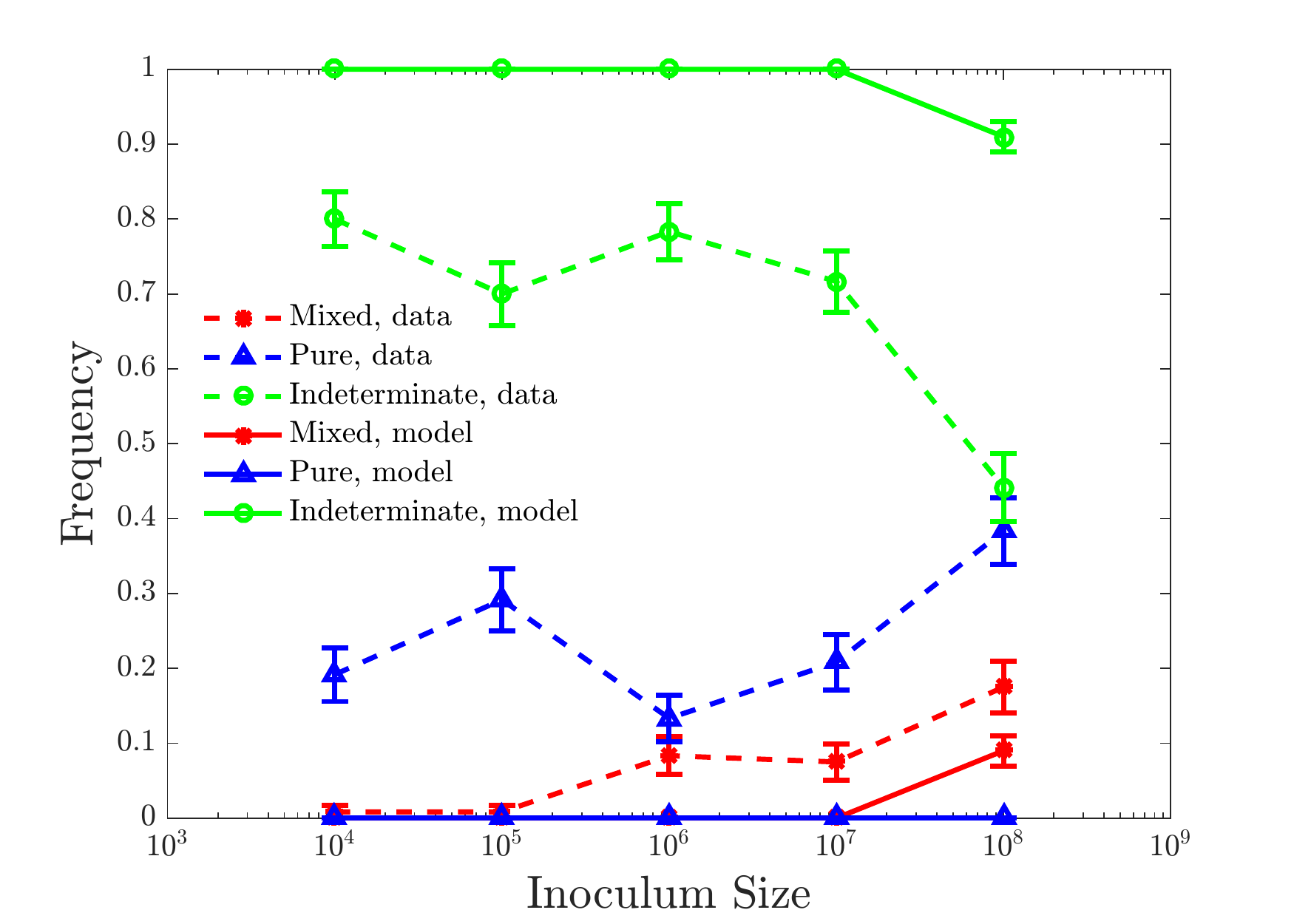}\includegraphics[width=.5\linewidth]{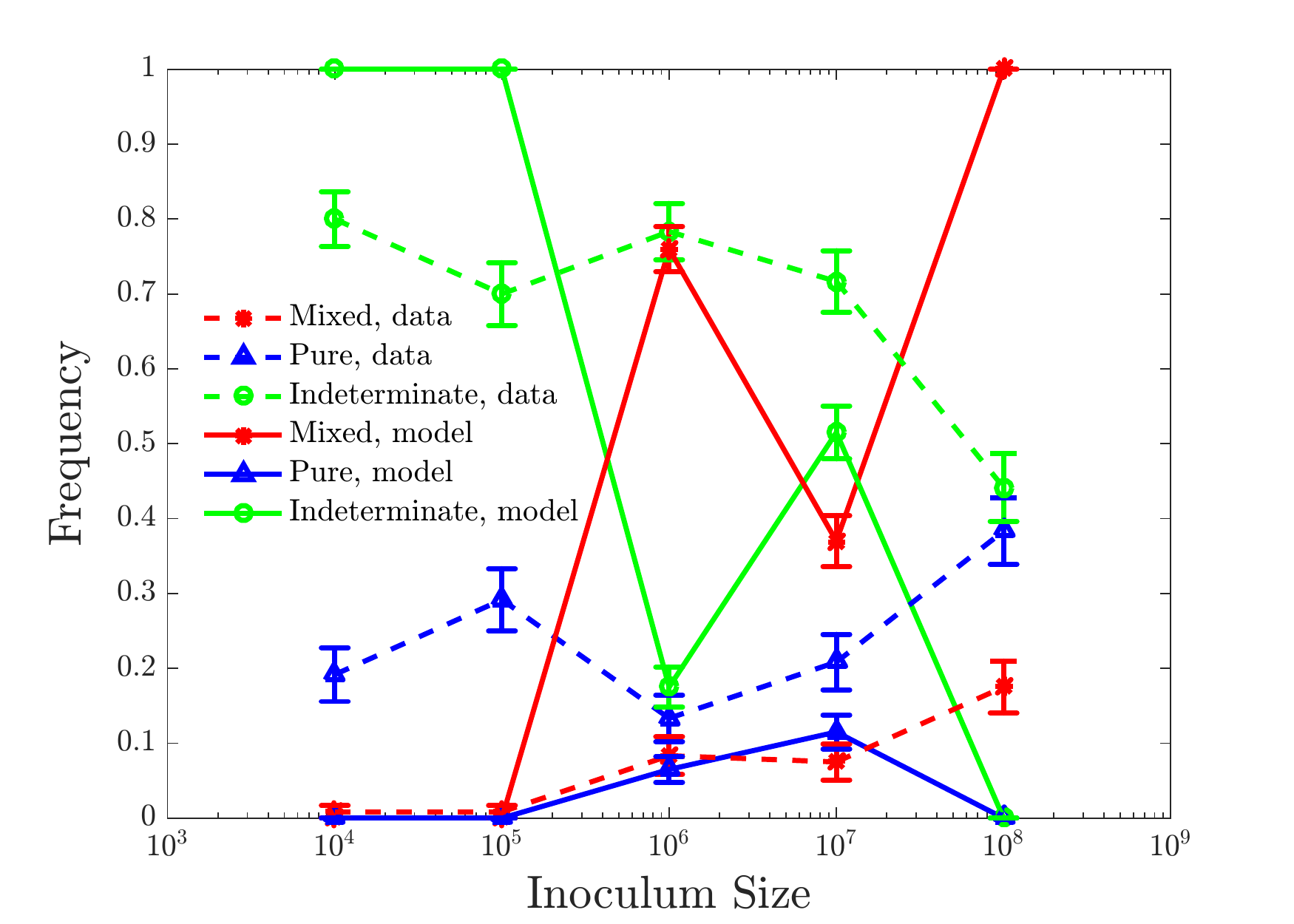} 
  \caption
  {{\bf Maximum likelihood fits of the colloquial model.}  We show two
    different local maxima in the parameter space with nearly
    equivalent likelihoods. Neither set of parameters provides good
    fits. In this and subsequent figures, error bars on model
    predictions are given by standard deviations of results from 200
    simulations. For the parameter values in the left panel
    ($\rho\approx1.7\cdot 10^{-5}$h$^{-1}$, $g_G=1.0$ h$^{-1}$,
    $r\approx6.1\cdot 10^{-6}$ h$^{-1}$ cells$^{-1}$,
    $\gamma\approx 240$ h$^{-1}$, $N_0\approx1.8\cdot10^{9}$ cells,
    $\alpha_j\approx4.5\cdot 10^{-5}$ h$^{-1}$,
    $d \approx 7.2\cdot 10^{-4}$ h$^{-1}$,
    $d_I \approx1.7\cdot 10^{-6}$ h$^{-1}$), infections do not
    establish until very large inoculums. For the right panel
    ($\rho\approx 1.7\cdot 10^{-5}$ h$^{-1}$, $g_G=1.0$ h$^{-1}$,
    $r\approx 4.1\cdot 10^{-8}$ h$^{-1}$ cells$^{-1}$,
    $\gamma\approx30.5$ h$^{-1}$, $N_0\approx1.2\cdot10^{8}$ cells,
    $\alpha_j\approx1.2\cdot 10^{-5}$ h$^{-1}$, $d = 0.01$ h$^{-1}$,
    $d_I = 0.018$ h$^{-1}$), the need to establish pure infections
    over the four orders of magnitude in the inoculum size leads to a
    large number of mixed infections as well.}
\label{fig:folkfit1}
\end{figure}

Two of the best quantitative fits of the colloquial model are shown in
Fig.~\ref{fig:folkfit1}. Some of the parameters of these fits are
physiologically unrealistic, but even this does not help: none of the
fits are good. The main difficulty seems to be that keeping the
fraction of pure / indeterminate infections nearly constant over four
orders of magnitude of the inoculum size, $S$, especially following a
mixed infection soon after inoculation, is impossible within this
independent action model. Indeed, the fit in the left panel keeps pure
infections at nearly zero frequency in order to have few mixed
infections. Similarly, in the right panel, which does a better job in
fitting the frequency of pure infections, the mixed infections rate
spikes to 100\% at high $S$. Note parenthetically that the
non-monotonicity of the mixed infection line in this figure is because
of the linearly increasing flux $j$ interacting with the immune
system. When $S=10^{4}$ and $10^{5}$, the flux is small, and
infections do not start. When $S=10^{6}$ and $10^{7}$, both of the
strains A and B have more than 100 bacteria of the {\em crossing}
phenotype in the blood even if none of the cells switches, resulting
in a mixed infection according to our definitions. At $S=10^7$,
switching starts happening often, resulting in a faster activation of
the immune response and clearance of the non-switched
strain. Finally, when $S=10^{8}$, both strains are in large enough
numbers to switch early on and at about the same time, resulting in
mixed infections again, but now in the {\em growing} phenotypes.

\begin{figure}[h]
\centering
    \includegraphics[width=.75\linewidth]{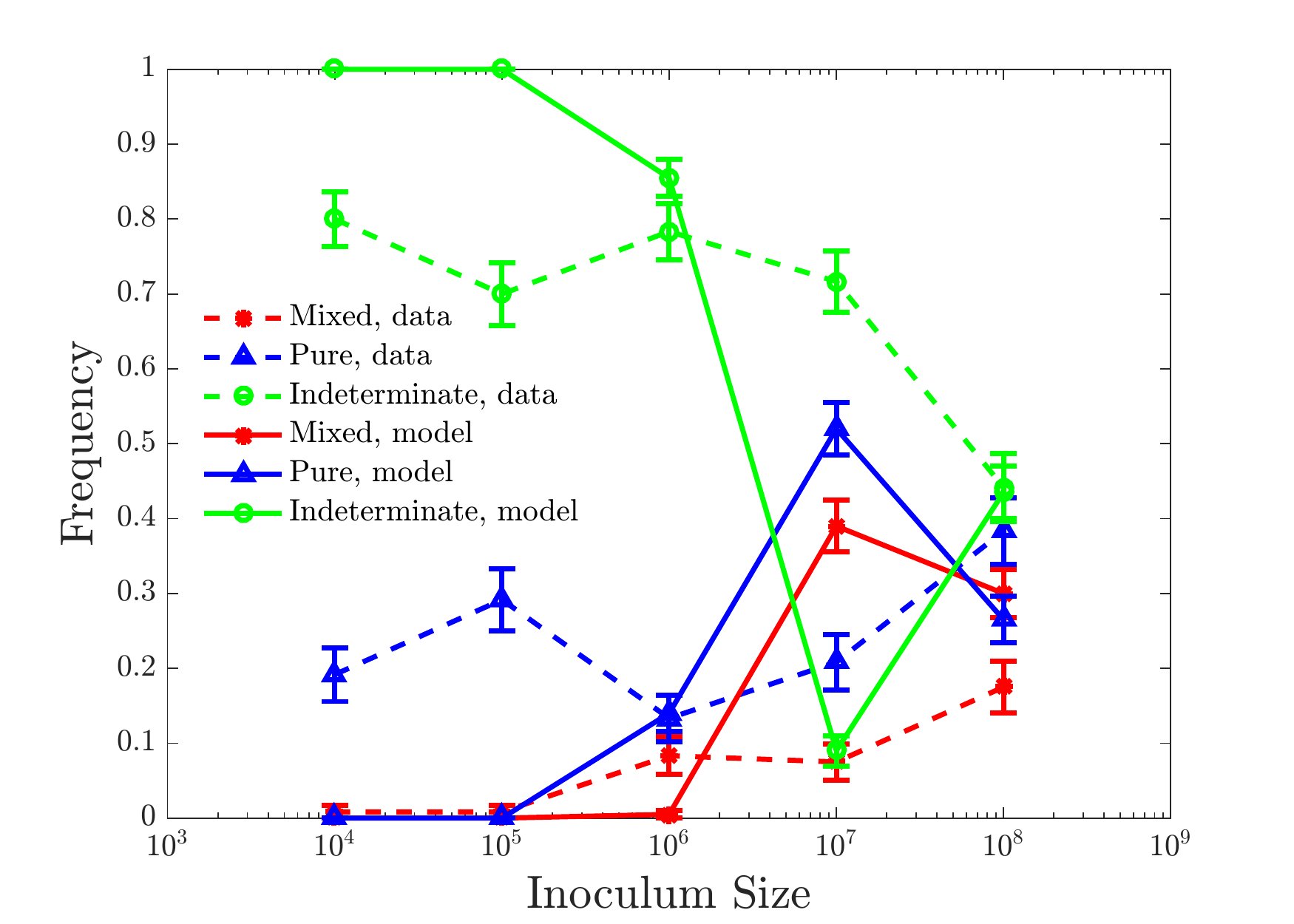} 
  \caption
  {{\bf Maximum likelihood fits of the colloquial model with the
      limited time duration of the bacterial flux from the nasal
      cavity to the bloodstream.} Natural bacterial cell death rate
    and immune cell death rate are set to zero.  The duration of the
    flux is a fixed value for all inoculum sizes, $t_j\approx 4.1$
    h. Other optimized parameters are: $\rho \approx3.5\cdot 10^{-5}$
    h$^{-1}$, $g_G\approx0.96$ h$^{-1}$,
    $\gamma\approx2.0$ h$^{-1}$,
    $r\approx2.1\cdot 10^{-7}$ h$^{-1}$ cells$^{-1}$,
    $N_0\approx1.0\cdot 10^{6}$ cells,
    $\alpha_j\approx1.4\times 10^{-5}$ h$^{-1}$.}
\label{fig:lfs1limt}
\end{figure}

One can modify the model to make it fit better. Since rats have
mucosal immunity \cite{barreau2004neonatal,lai2004glutamine,
  lee2004antimicrobial,lysenko2005role,margolis2010ecology}, one can
hope that bacteria in the nasopharynx will be eventually cleared as
well. The simplest way of modeling this is to say that the flux from
the nasopharynx into the bloodstream has a finite duration $t_j$,
which itself is an unknown variable that needs to be fitted. Further,
we notice that the natural (not caused by the immune system) death
rate of bacterial cells and the death rate of immune cells in the fits
in Fig.~\ref{fig:folkfit1} are very small. Hence, to not simplify
fitting with the introduction of the new time parameter, we set both
of these parameters to zero (we verified that the fits do not improve
dramatically when this condition is relaxed). The optimized fit for
this model is shown in Fig.~\ref{fig:lfs1limt}.  The fit is clearly
better than in Fig.~\ref{fig:folkfit1}, but it is still poor: to have
pure infections at medium/high inoculums, the independent action
hypothesis still requires no (or indeterminate) infections at small
inoculums.

The key problem of the colloquial model is the experimentally observed
weak dependence of the fractions of various infection types on the
inoculum size. In other words, in these simple models, independent
action means that the number of cells that attempt C$\to$G switching
in the blood scales with $S$. Thus the time to the first such switch
would scale as $1/S$. If the switch happens in the bulk of the 54 h
experiment duration, an infection is established. Thus it is very hard
to devise an independent action model that would have a non-negligible
number of switches over 54 hrs at small inoculums, $S=10^4$, and yet
would not have switches happening 100\% of the time at large
inoculums, $S=10^8$. The model must be modified so that the mean time
to the first switch decreases slower than $1/S$.

Interestingly, there is a straightforward biologically realistic
modification of the model that achieves this. In many cases, the
process of bacterial phenotypic switching is not determined by a
one-step chemical reaction, but proceeds through a series of roughly
equally slow steps.  For example, the switching of {\em E.~coli} to
express PAP genes and become virulent
\cite{hernday2002self,hernday2003mechanism} can be modeled as a
four-step reaction \cite{pelevs2006reduction}. Such $n$-step
activation ensures that the probability distribution of time to the
complete switch in an individual bacterium goes as $\propto t^{n-1}$
for small $t$ \cite{doan,cheng2013large}. Then for $\propto S$
bacteria, the expected time till the first of them switches is
governed by the Weibull distribution, resulting in
$St^{n-1}\propto 1$, and $t\propto 1/S^{1/(n-1)}$
\cite{cheng2013large}. In other words, the time to the first bacterium
in a large population switching scales sublinearly with the inverse
population size, offering a potential opportunity to explain the weak
dependence on the inoculum size. 

\begin{figure}[h]
\centering
    \includegraphics[width=.75\linewidth]{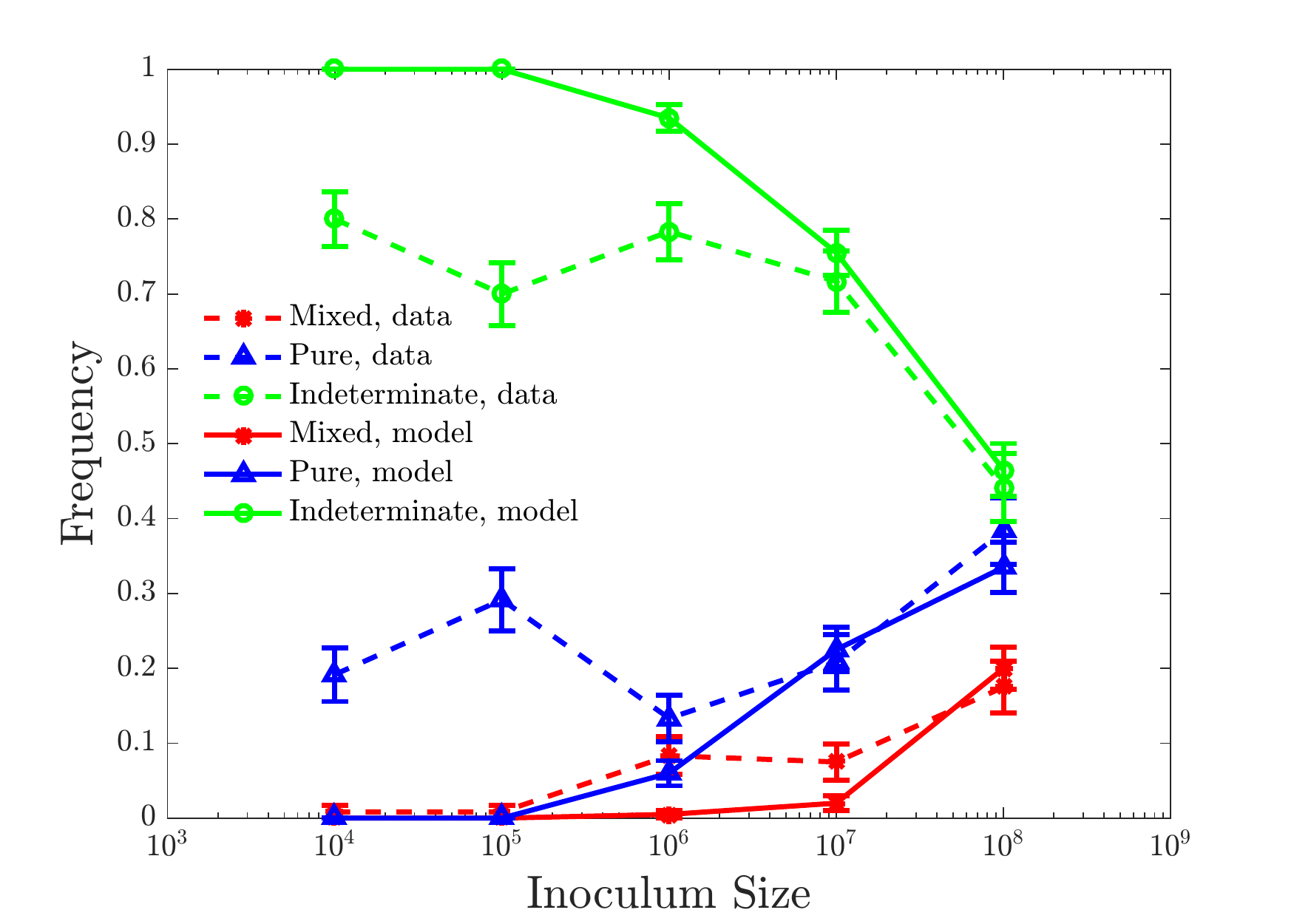} 
  \caption
  {{\bf Maximum likelihood fits for the colloquial model with the
      limited bacterial flux duration and three-step switching.}
    Natural bacterial cell death rate and immune cell death rate are
    set to zero. The duration of the flux is a fixed value for all
    inoculum sizes, $t_j\approx 4.1$ h.  Optimized parameter values
    are $\rho\approx0.0014$ h$^{-1}$, $g_G\approx1.1$ h$^{-1}$,
    $\gamma\approx9.7$ h$^{-1}$,
    $r\approx1.1\cdot 10^{-7}$ h$^{-1}$ cells$^{-1}$,
    $N_0\approx1.0\cdot 10^{9}$ cells, 
    $\alpha_j\approx3.4\cdot 10^{-4}$ h$^{-1}$.}
\label{fig:lfs3t4}
\end{figure}

We implement and optimized this model in numerical simulations by
introducing a series of phenotypic transitions
C$\to$G$_1\to\dots\to$G$_n$, where only the last state, G$_n$, grows
fast, and the rest of the states share the growth/death rates with
C. Random switching between the subsequent states was again governed
by the Poisson dynamics, as in Eq.~(\ref{eq:ppoiss}). We explored
$n=2,3,4$. Fig.~\ref{fig:lfs3t4} shows results of the optimization,
where $n=3$ (empirically best choice), and all transition rates in the chain
C$\to$G$_1\to$G$_2\to$G$_3$ were the same (which results in the most
sublinear dependence of the switch time on $S$). Further, since in
this model switching takes extended time, the first bacteria to cross
over to blood from the nasopharynx will be the ones switching, and
hence it makes little difference for the switching statistics if the
flux has a limited duration. At the same time, stopping of the flux
into the blood results in a lower concentration of the non-switched
strain, making it easier to develop pure infections at 54
hrs. Therefore, we inherit the value $t_j\approx4.1$ hrs from the
1-step model. Clearly, the quality of fit improves dramatically
compared to the 1-step model, and yet the fits are still far from
perfect: mixed and pure infections go hand-in-hand, and to have no
mixed infections at $S=10^4$ requires having no infections at all at
this inoculum. This illustrates a fundamental problem of the
multi-step switching mechanism: while the time to a switch, indeed,
scales sublinearly with $1/S$, the standard deviation of this time
falls off very quickly, making the switching nearly deterministic
\cite{cheng2013large}. Thus both strains switch at about the same
time, and typically either both develop into an infection (mixed
outcome) or none does (indeterminate outcome). 

In summary, the independent action model, even augmented by multi-step
switching and finite bacterial flux duration, seems to be incapable of
explaining the observed experimental data.

\subsection{Beyond the independent action model}

The independent action hypothesis is implemented in our model, in
part, by an assumption that the bacterial flux from the nasopharynx to
blood is proportional to the inoculum size, $S$. We consider multiple
extensions of the colloquial model that break this assumption.

\begin{figure}[h]
\centering
    \includegraphics[width=.75\linewidth]{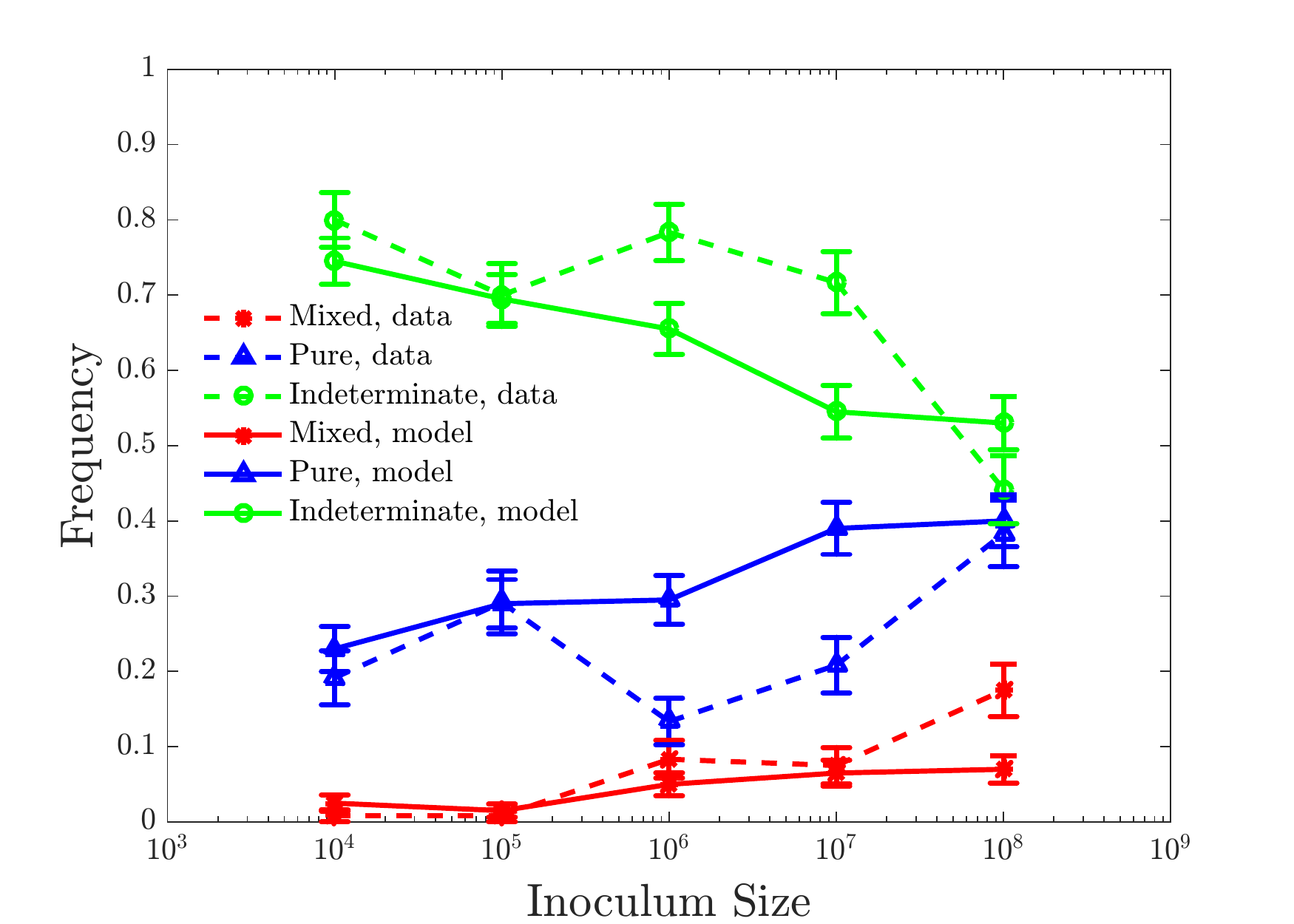} 
  \caption
  {{\bf Maximum likelihood fits for the non-independent action model
      with the $S$-dependent flux duration.} The fitted model has
    $j=\alpha_j={\rm const}$, and $t_j=\alpha_t S^{b_t}$. This model provides
    much better fits than all of the variants of the independent
    action model we have tested. The optimized parameters are
    $\rho\approx 6.1\cdot 10^{-6}$ h$^{-1}$, $g_G\approx0.55$ h$^{-1}$,
    $N_0\approx1.0\cdot10^{7}$  cells, 
    $\gamma\approx2.0$ h$^{-1}$,
    $r\approx 2.3\cdot 10^{-6}$ h$^{-1}$ cells$^{-1}$,  
    $\alpha_t\approx 0.5$ h$^{-1}$, $b_t \approx
    0.1$,  $\alpha_j \approx 3.3\cdot 10^3$ h$^{-1}$. }
\label{fig:fixjnlt}
\end{figure}

First, we tried the model where the flux is independent of $S$,
$j=\alpha_j$, which must be optimized. However, then the flux duration
scales nonlinearly with $S$, $t_j=\alpha_t S^{b_t}$. The logic behind
this model is that there might be purely physical constraints on how
many bacteria can cross the tissues between the two body compartment
per unit time, and this bandwidth can be saturated even at moderate
inoculums. At the same time, it could take the mucosal immunity a
longer time to clear a larger inoculum. We retain the 1-step switching
model, because the higher variability of the switching time within
this model allows for easier establishment of pure infections.  The
maximum likelihood results are shown in Fig.~\ref{fig:fixjnlt}. While
imperfect, the fits are surprisingly good, able to sustain pure,
mixed, and indeterminate infections over the entire range of
$S$. However, $b_t\approx 0.1$ is very small, so that the duration of
the bacterial flux is maximally $\sim 3$ hours, which makes it hard to
imagine physiological mechanisms that would create it.

\begin{figure}[t]
\centering
    \includegraphics[width=.75\linewidth]{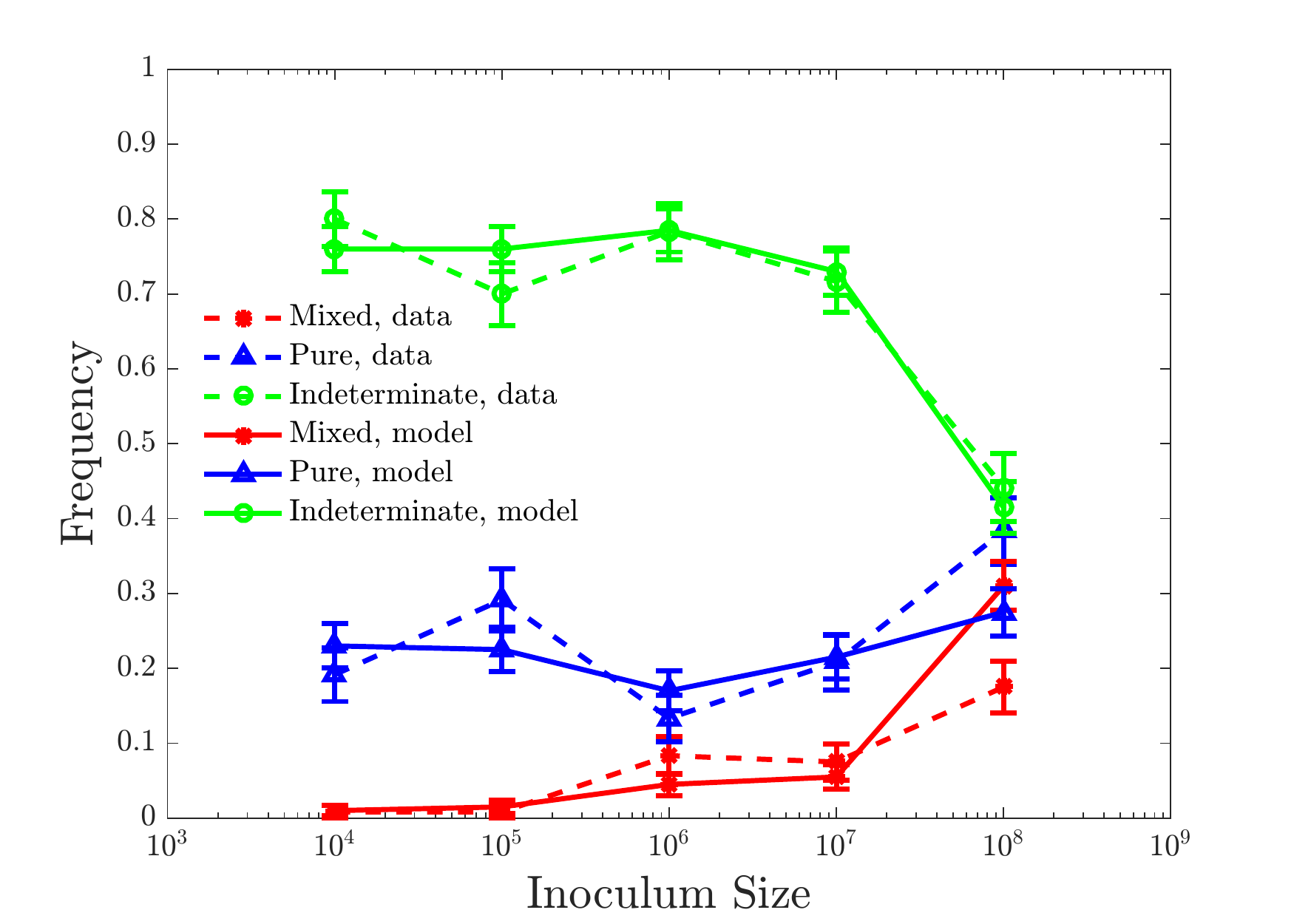} 
  \caption
  {{\bf Maximum likelihood fits for the non-independent action model
      with the sublinear dependence of the magnitude of the bacterial
      flux on $S$.} We use $j=\alpha_jS^{b_j}$ with a fixed $t_j$. The
    optimized parameters are: $\rho\approx5.7\cdot 10^{-6}$ h$^{-1}$,
    $g_G\approx1.0$ h$^{-1}$,
    $\gamma\approx3.2$ h$^{-1}$,
    $r\approx 3.1\cdot 10^{-6}$ h$^{-1}$ cells$^{-1}$,
    $N_0\approx1.0\cdot 10^{6}$ cells, $\alpha_j\approx7$ h$^{-1}$,
    $b_j \approx 0.37$, and finally $t_j \approx 34.0$ h, which is
    essentially equivalent to saying that the bacterial flux is
    temporally unconstrained. }
\label{fig:nljfixt}
\end{figure}

Another way to model non-independent action is the model where the
duration of the bacterial flux $t_j$ is fixed and independent of $S$,
but $j=\alpha_j S^{b_j}$. Optimized dynamics for this model is shown
in Fig.~\ref{fig:nljfixt}, providing clearly the best fit to the
experimental data of the ones we have seen so far. Interestingly, the
fitted values of the parameters in this model are biologically
realistic, resulting, for example, in bacterial fluxes of
$10^2\sim10^3$ cells$/$h, and $t_j\approx 34$ hrs, longer than 1 day.

\section{Discussion}

In this investigation, we built mathematical models of the early {\em
  Hib} stages of infections started by a culture with two equally
invasive strains.  The models have to account for the following broad
observations: (1) both strains are present in the blood soon after
inoculation into nasal passages; (2) both strains are either cleared
from the blood within 54 hrs post inoculation, or are equally likely
to be responsible for monoclonal bacteremias; and (3) the weak
dependence of the likelihood of invasive infections on the inoculum
size, i.~e., a few-fold difference in the likelihood of bacteremia for
a $10^4$-fold change in the size of the inoculum. The colloquial
hypothesis tries to explain these effects as variations on the theme
of independent action of bacteria in establishing infections.  In such
explanations, one commonly augments the independent action by
stochasticity of phenotypic transitions and by interactions with the
immune system.  In this interpretation, both bacterial clones
inoculated into the nasal passages cross into the bloodstream, at
which point one individual randomly switches into a more invasive
phenotype and activates the immune system. The immune system then
clears the non-switched strain, resulting in monoclonal infections.

We analyzed this colloquial hypothesis quantitatively, starting with
its simplest realization. Further, we considered additional variants
that involved more complex (and hence statistically different)
switching between the crossing and the growing phenotypes, or effects
of mucosal immunity, which can clear the nasopharyngeal infection and
stop bacterial flux into the bloodstream a few hours into the
experiment. None of these modifications proved sufficient to explain
the experimental data, and, in particular, to account for the weak
dependence of the likelihood of the infections on the inoculum size.
In contrast, when we forwent the independent action assumption and
allowed the flux of bacteria into the bloodstream to depend
sublinearly on the inoculum size, the fits to the data became very
good. Thus our analysis suggests that the hypothesis of independent
action may be violated in the case of early establishment of bacterial
infections. Note that classic investigations of independent action
\cite{meynell1957some} tested the hypothesis against {\em synergistic}
effects. Here we argue that the non-independent action effects are
redundant---the probability of a single bacterium to establish an
infection decreases with the inoculum size. This is surprising, and
certainly goes against ideas in the quorum sensing literature
\cite{Waters:2005dt}, where an infection is established
synergistically when the number of bacteria crosses a certain minium
density threshold.

Our best model suggests that the flux of bacteria from the
nasopharyngeal inoculation to the bloodstream scales as the inoculum
size to the $\sim 0.37$ power. The data that we have been able to find
in the literature is not sufficient to provide an empirical basis for
the mechanism of this scaling: the physical structure of the animal
tissues and bacterial colonies, the fluid dynamics of the bacterial
culture in the nasopharyngeal cavity, interactions of bacteria with
the immune system, or interactions of bacteria among themselves could
all play a role. The closeness of the exponent to $1/3$ is also
interesting, suggesting that maybe a certain modification of the
3-step stochastic switching model, similar to that studied in
Fig.~\ref{fig:lfs3t4}, could play a role as well.

While informative at some level, negative results alone rarely appear
in publications nowadays \cite{Fanelli:2011hj}. Beyond obvious social
pressures, there are functional reasons for this as well: one can
never be sure that the negative result is meaningful, rather than due
to not trying hard to find a positive agreement between a hypothesis
and data.  For example, in our study, we cannot be sure that we have
explored the parameter space well enough, and that we have tried all
simple, reasonable modifications to the original colloquial model to
argue that the independent action theory cannot explain the data. We
can only say with certainty that {\em we have not been able to
  reconcile} the independent action theory with the data. It will
require additional experimental and theoretical investigations to
understand if and under which conditions the independent action
hypothesis is, indeed, violated in early infections. The model
proposed here would suggest that the bacterial concentration in the
blood soon after inoculation should scale sublinearly with the
incoculum size, which is relatively straight forward to check
experimentally. Further, repeating the experiments in immune
compromised rats, with no innate immunity, should result in no pure
infections even at small inoculums. Moreover, due to the sublinear
dependence, the probability of transmission of the disease among
individuals should depend weakly on the strength of the infection in
an infected host, which is bound to have public health implications,
and is also experimentally testable. We hope that our study will spur
such future investigations, in {\em Hib} and in other infections.

\section*{Acknowledgements}
This work was partially supported by the James S.~McDonnell Foundation
grant No.~220020321 (IN), by the National Science Foundation grant
No.~PoLS-1410978 (IN), and by National Institutes of Health grant
No.~GM098175 (BL). We thank Rustom Antia and Richard Moxon for
illuminating discussions.

\section*{References}

\providecommand{\newblock}{}

\end{document}